# Butterfly Effect Confirmed in Global AI Weather Models: Evidence from Tropical Cyclone Forecasting

Jeremy Cheuk-Hin Leung[1,2,3#], Daosheng Xu[1,2#], Weiye Yu[1,2], Shaojing Zhang[1,4], Xiaodong Zeng[5], Gaozhen Nie[6], Jie Feng[7], Jingchen Pu[7], Yi Li[1,2,8*], Kaijun Ren[1,2], Qingcun Zeng[5] & Banglin Zhang[1,2,3,4,8*]

**Affiliations:**

[1] College of Meteorology and Oceanography, National University of Defense Technology; Changsha, China

[2] Laboratory of Atmospheric Environmental Monitoring and Early Warning for Low-Altitude Economy, National University of Defense Technology, Changsha, China

[3] Hunan Institute of Advanced Technology, Changsha, China

[4] College of Atmospheric Sciences, Lanzhou University, Lanzhou, China

[5] Institute of Atmospheric Physics, Chinese Academy of Sciences, Beijing, China

[6] National Meteorological Centre, China Meteorological Administration, Beijing, China

[7] Department of Atmospheric and Oceanic Sciences and Institute of Atmospheric Sciences, Fudan University, Shanghai, China

[8] Key Laboratory of High Impact Weather (special), China Meteorological Administration, Changsha, China

# These authors contributed equally to this work.

* Corresponding authors. Email: zhangbanglin24@nudt.edu.cn; liyiqxxy@163.com

**Abstract:**

A paradox recently emerged in artificial intelligence (AI) weather prediction research. While some claim AI weather models cannot simulate atmospheric butterfly effect, this conflicts with AI models' limited predictability and advances in AI ensemble forecasting. This study demonstrates via counterexamples that the butterfly effect does exist in AI weather predictions. For Super Typhoon Khanun, AI predictions are constrained by a double-attractor system. Minor initial perturbations confined to two regions trigger state transitions between two local attractors, causing a 1006-km difference in the predicted storm position on Day 7. This behavior is consistent with numerical weather prediction models and observed in ~12% of tropical cyclones in the past 5 years. These findings verify AI's ability to capture atmospheric chaos and provide the physical basis for AI ensemble forecasting.

**Main Text:**

The Earth's atmosphere is a prototypical chaotic system, whose evolution is sensitive to minor discrepancies in its initial state (*1*, *2*). This chaotic nature allows tiny initial errors to grow exponentially over forecast lead times, imposing a fundamental theoretical limit on weather predictability. This is often referred to as the predictability barrier, or more colloquially, the butterfly effect (*3*), and serves as the fundamental physical principle for ensemble forecasting.

By simulating atmospheric chaoticity, ensemble forecasting generates probabilistic outputs that quantify prediction uncertainty, enhancing the reliability of numerical weather prediction (NWP) (*4–6*). Probabilistic ensemble forecasts have long served as an essential foundation for routine meteorological services, disaster mitigation, and government decision-making. Since the proposal of the ensemble forecast framework in the 1990s (*7–10*), it has been widely applied in operational forecasting centers around the world (*11–18*).

Until recently, the rapid breakthroughs of artificial intelligence (AI)-driven weather prediction have opened up a new direction to solve the weather prediction problems (*19–23*). AI weather models not only achieve prediction accuracy comparable to state-of-the-art NWP systems but also exhibit superior computational efficiency once the model is trained. Their low computational cost enables massive ensemble simulations, which directly overcomes a major bottleneck of traditional NWP-based ensemble forecasting. This advantage addresses the time-sensitive demands of real-time operational weather forecasts and provides new possibilities for large-scale ensemble prediction (*24–27*).

However, a paradox has recently arisen regarding the chaotic characteristics in AI weather models. On one hand, the ensemble techniques have been successfully applied to AI forecasting systems and have effectively improved AI prediction performance (*24*, *27–30*), suggesting that existing global AI weather models can capture atmospheric chaotic behavior. On the other hand, recent publications have questioned the ability of these data-driven models to reproduce the atmospheric butterfly effect. They argued that AI models fail to replicate the rapid growth of small-amplitude initial perturbations, a defining feature of the butterfly effect observed in conventional NWP models (*31*, *32*).

Whether the butterfly effect exists in AI weather models remains an open and controversial question. From a theoretical perspective, chaos theory suggests that the butterfly effect sets an upper limit on weather predictability. In this context, the absence of butterfly effect in AI weather models would incorrectly imply unlimited forecast predictability (*33*, *34*). From a practical perspective, if AI weather models truly failed to capture atmospheric chaoticity, ensemble forecasting would theoretically lose its physical basis when applied to AI predictions, which contradicts the well-documented effectiveness of AI ensemble experiments (*19–22*).

Current global AI weather models are trained on NWP-derived reanalysis datasets (*35*), and it is based on these NWP models that the chaotic nature and butterfly effect of atmospheric systems were originally identified and validated (*36*). In principle, well-trained AI models should therefore be able to learn and reproduce such chaotic dynamics. This inference is supported by previous studies confirming that machine learning models can effectively learn chaotic behaviors of the atmosphere and other complex systems from training data (*37–40*). Given that global AI weather models exhibit forecasting skills comparable to NWP models, we hypothesize that the butterfly effect does exist in global AI weather models, but it may manifest in a manner distinct from that of traditional NWP models, which gives rise to the "paradox".

Clarifying the existence of the butterfly effect in AI weather models is a key issue for evaluating the physical rationality of data-driven forecasting systems and provides a theoretical basis for the future development of AI ensemble prediction. Thus, this study aims to objectively explore the existence of the butterfly effect in current global AI weather models and its occurrence conditions. The following analysis is conducted specifically in the context of tropical cyclone (TC) track forecasting, where the sensitivity of prediction results to perturbations can be observed intuitively.

**A Counterexample: Super Typhoon Khanun (2023)**

In the following, we present a counterexample, Super Typhoon Khanun (2023), to demonstrate that AI TC trajectory forecasts can be sensitive to initial conditions, a fundamental feature of the butterfly effect.

Super Typhoon Khanun affected the western North Pacific (WNP) in 2023. It was distinguished by its unusual trajectory, experiencing two sharp turns on 3rd and 7th August, respectively (Fig. 1A, Text S1). Due to its extremely unusual trajectory, combined with strong intensity and a long lifetime, Khanun caused substantial damage in East Asia, including Japan, China, North Korea, and South Korea (*41*, *42*).

Particularly, the first sharp turn of Khanun was not accurately predicted by both global NWP and AI weather models up to four days in advance (*42–44*). When initialized from 0000UTC 28th to 0000UTC 30th July 2023, Pangu-Weather, one of the AI weather models renowned for skillful TC track predictions (*19*), consistently predicted Khanun to move northwestward and make landfall in East China (Fig. 1B). However, starting from 1200UTC 30th July, Pangu-Weather abruptly revised its forecast, predicting Khanun to make a sharp right turn near the Ryukyu Islands and subsequently head eastward (Fig. 1C). The sharp decline in Khanun's track forecast error at 1200UTC 30th July verifies this rapid change in predicted track that arose from changes in initial conditions within a 12-hour interval of initialization times (Fig. 1D).

The rapid change in forecast outcomes between two close initialization times somehow indicates that Pangu-Weather's predictions of Khanun's trajectory are sensitive to small variations in initial conditions. In most cases, no matter for NWP or AI weather models, TC trajectory forecasts typically undergo gradual adjustments as initialization times progress. However, the Khanun case is a clear exception to this typical pattern. This rapid adjustment of forecast results resembles a state jumping from one local attractor to another within a 12-hour interval of initialization times.

We found similar rapid adjustments of forecast results in 14 WNP TCs during 2020–2024, implying that Khanun is not an exceptional case exhibiting such behavior (Text S2, Fig. S1). Furthermore, NWP models also yield similar rapid adjustment in Khanun's track prediction (*42*, *43*) (Fig. S2), implying that this sensitivity of AI weather predictions to initial conditions is equivalent to that in NWP models.

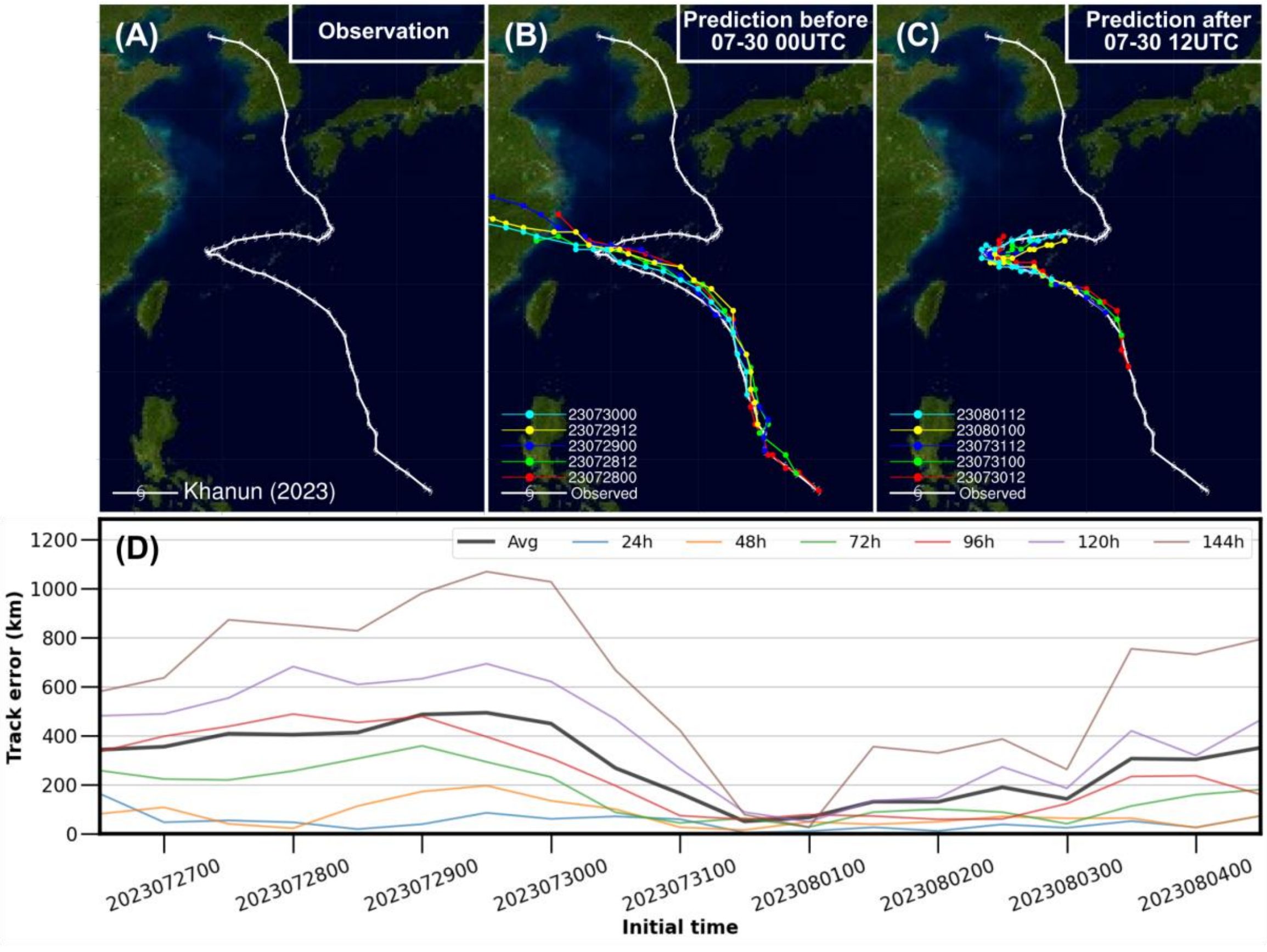

**Fig. 1. Rapid adjustment of Pangu-Weather's forecast results for the abrupt right turn of Super Typhoon Khanun. (A)** Observed trajectory of Super Typhoon Khanun. **(B)** Predictions of Pangu-Weather on Khanun's trajectory, initialized from 0000 UTC 28th to 0000 UTC 30th July 2023. **(C)** Same as **(B)** except for predictions initialized from 1200 UTC 30th July to 1200 UTC 1st August 2023. **(D)** Pangu-Weather's forecast error (unit: km) of Khanun's trajectory against initialization time. Different colors denote forecast error at different forecast lead times, with their averages denoted by the black curve.

## Double Attractors in AI Probabilistic Prediction Experiment

To further verify the presence of butterfly effect in AI weather models, we conduct a probabilistic prediction experiment, aiming to examine the sensitivity of Pangu-Weather's predictions to initial perturbations. The fast physics-constrained perturbation generator (*27*) (see Methods) was adopted to generate initial ensemble perturbations at 1200 UTC 30$^{th}$ July, the initialization time when Pangu-Weather's forecast error sharply reduces (Fig. 1A).

The experiment produces a large spread in Khanun's trajectory forecast result, characterized by two distinct clusters of predicted tracks. Among the 528 ensemble members, approximately 58% of them correctly predicted Khanun's right turn after passing through the Ryukyu Islands (referred to as the right-turn cluster). Another 24% predicted Khanun to continue moving northwestward and make landfall in China (referred to as the landfall cluster). A minor proportion (18%) predicted northward track or other moving directions (Fig. 2A).

A notable feature is that the large forecast spread is neither randomly nor evenly distributed. The two clusters of predicted Khanun's trajectories, each oriented toward a specific direction, subsequently form a double-probability center pattern beyond 120-hour lead time. During the first 120 hours of forecast, Khanun's positions are generally concentrated in a single probability center, corresponding to Khanun's movement from its initial position to the northwest of the Ryukyu Islands. As the forecast progresses, the probability center then splits into two (Fig. 2C), corresponding to the right-turn and landfall clusters, respectively.

This non-random ensemble spread obeys the bounded behavior of chaotic systems, suggesting that the sensitivity of Pangu-Weather to initial perturbations is somehow characterized by two local attractors. One attractor governs Khanun's abrupt right turn, while the other governs its northwestward movement. This feature is consistent with that simulated in NWP ensemble prediction by ECMWF, which also exhibits a clear double-attractor system (Figs. 2B, 2D, S3).

This chaotic behavior in Pangu-Weather's predictions is further supported by quantitative measures. Both the ensemble mean and standard deviation of prediction error exhibit a nonlinear growth driven by discrepancies in initial conditions (Fig. 2E). Particularly, the growth accelerates after 120-hour lead time, which directly corresponds to the appearance of a double-probability center pattern (Fig. 2C). Further, The nonlinear local Lyapunov exponent (NLLE) of the ensemble result, a more classical way to quantify the degree of chaos (*45–47*), decreases monotonically and quickly converges to approximately 0.7 (Fig. 2F). The positive NLLE value verifies the presence of chaotic behavior in the model.

The above ensemble experiment confirms the fact that Pangu-Weather's prediction on Khanun's right turn is sensitive to initial perturbations, a fundamental feature of the butterfly effect.

More importantly, the forecast exhibits a distinct double-probability center pattern, which resembles the presence of two local attractors governing the system.

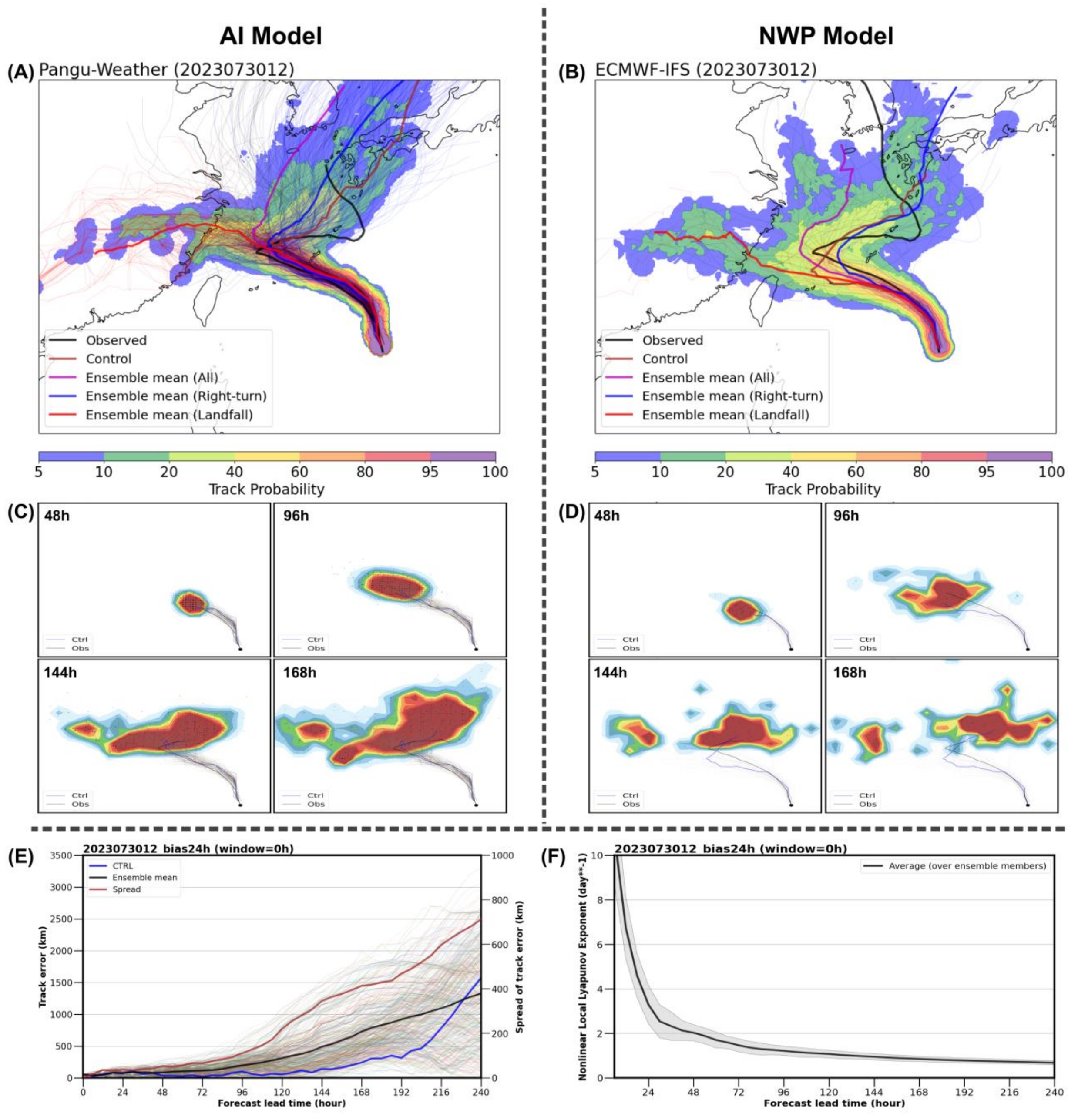

**Fig. 2. Double-probability center pattern in the spread of Pangu-Weather's probabilistic prediction on Khanun's abrupt right turn. (A)** Track probability (shading, unit: %) predicted by 528 ensemble members (thin curves) of Pangu-Weather on Khanun's abrupt right turn, with initialized at 1200UTC 30th July 2023. The black curve denotes the observed Khanun's trajectory. The red, blue, and purple curves respectively denote the averages of tracks moving northwestward (red, 126

members), experiencing a right turn (blue, 306 members), and toward northward or other directions (purple, 96 members). The brown curve denotes the results of control run. The initial perturbations were generated by the fast physics-constrained perturbation generator (*27*). **(B)** Same as **(A)**, except for ensemble results of ECMWF-IFS. **(C)** Probability distribution of Khanun's predicted positions by Pangu-Weather across different forecast lead times: 48, 96, 144, and 168 hours. **(D)** Same as **(C)**, except for ensemble results of ECMWF-IFS. **(E)** Prediction error (thin curves, unit: km) of each ensemble member of Pangu-Weather initialized at 1200UTC 30$^{th}$ July 2023, and the corresponding ensemble mean (black thick curve, unit: km) and ensemble spread (red thick curve, unit: km). **(F)** The NLLE of the Pangu-Weather probabilistic prediction initialized at 1200UTC 30$^{th}$ July 2023, with the grey shading denoting the ±1 standard deviation of NLLE.

## Factors triggering the state transition

The above sections present a counterexample that demonstrates the presence of the butterfly effect in Pangu-Weather. It is essential to further explore the key factors that determine which local attractor each ensemble member eventually follows. This can be identified by comparing the initial conditions between the right-turn and landfall clusters in Fig. 2A. Through sensitivity experiments, we find that initial perturbations over two regions (Region A and B in Fig. S4) are the dominant factor driving the state transition.

For the unperturbed simulation initialized at 1200 UTC 30 July, Khanun is predicted to undergo an abrupt right turn on Day 4. Surprisingly, initial perturbations confined to Region A and B (namely Exp-AB, Fig. S4) are sufficient to suppress Khanun's abrupt right-turn, driving it toward northwestward and eventual landfall over China (Fig. 3A). Further diagnostic analysis suggests that the perturbed simulation produces a stronger western Pacific subtropical high (WPSH) to the north of Khanun and weaker southwesterly monsoon wind (Figs. 3D, 3E). The combined effect weakens the prevailing southwesterly steering flow, which suppresses Khanun's right turn. In addition, Khanun moves faster in the perturbed run before its right turn, due to the stronger easterly component of steering flow throughout the troposphere (Fig. 3F). This drives Khanun away from the southwesterly monsoon channel on Day 4, hence being less influenced by southwesterly monsoon wind and sustaining its northeastward movement. Introducing initial perturbations solely in these two key regions causes 1006 km differences in Khanun's 7-day predicted position.

Interestingly, introducing perturbations of either region alone is insufficient to trigger this track transition. When initial perturbations are imposed only over Region A (namely Exp-A),

Khanun exhibits a slightly more northwestward displacement, but eventually recurves northeastward toward Japan instead of making landfall. Meanwhile, perturbations limited to Region B (namely Exp-B) produce forecast tracks almost the same as those of the unperturbed run (Figs. 3B, 3C). These indicate that perturbations of both Region A and B are necessary conditions for triggering the prediction state transition.

Further analysis reveals that perturbations of Region A primarily accelerate Khanun's northwestward motion before its right-turn, whereas perturbations of Region B drive Khanun's landfall after 96-hour lead time. These are evident in the different environmental steering flows simulated in Exp-A and Exp-B. Pangu-Weather produces a stronger easterly component of tropospheric steering flow only in Exp-A (Fig. 3B, 3G), which accelerates Khanun's northwestward-moving speed before its right turn. By contrast, Exp-B exhibits negligible changes in Khanun's moving speed (Fig. 3C). However, the absence of perturbations in Region B respectively weakens and strengthens the mid-high tropospheric easterly and southerly steering flow (Fig. 3H and 3I), which drives Khanun head northeastward instead of making landfall. As a result, Khanun's acceleration induced solely by perturbations over Region A is insufficient to force Khanun to make landfall. Only the combined effects of Khanun's acceleration and modulated steering flow can trigger the track transition from one local attractor to the other.

These findings demonstrate that data-driven AI weather models, like traditional NWP models, are indeed able to capture the chaotic dynamics of atmospheric systems.

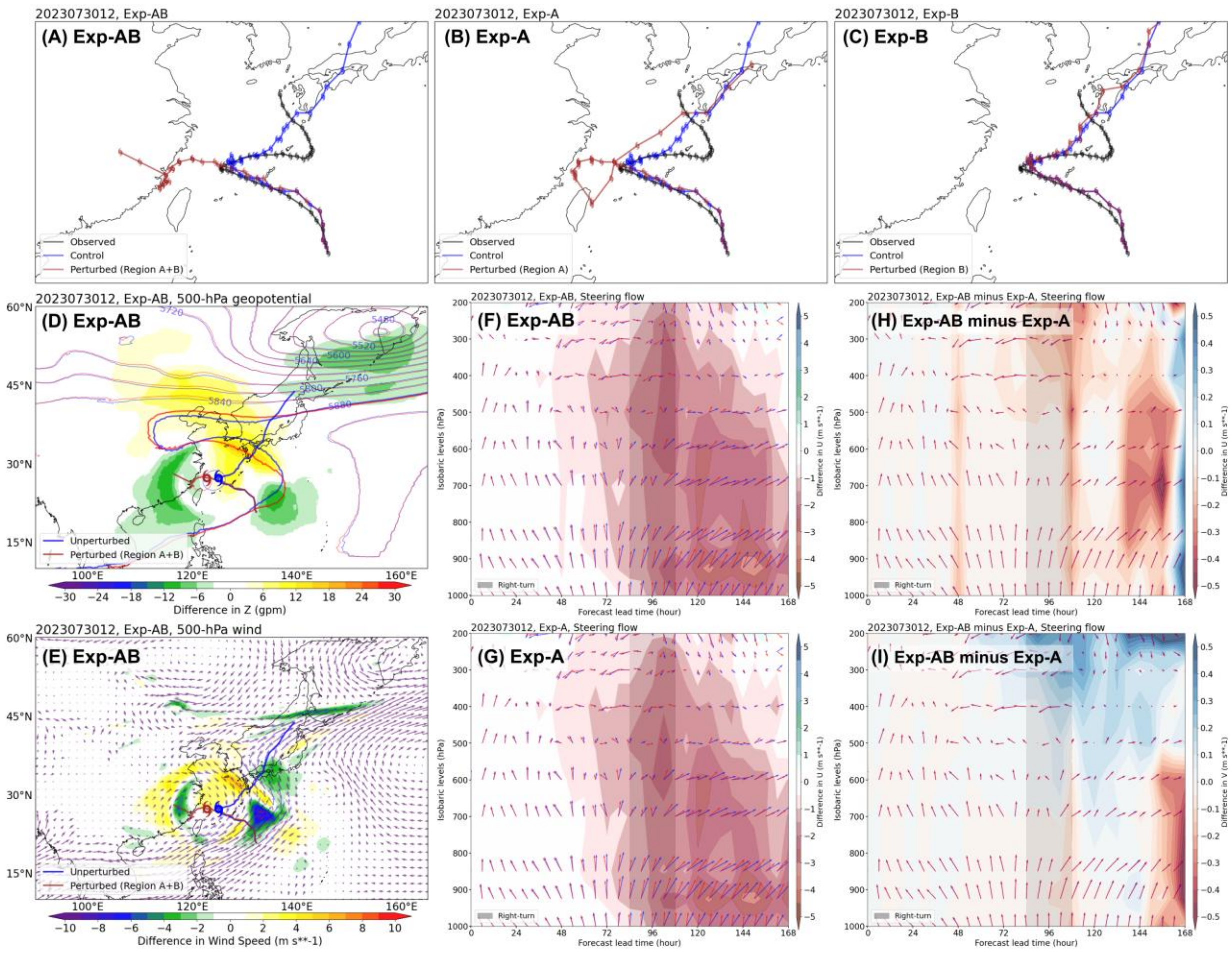

**Fig. 3. Sensitivity experiments of initial perturbations on Khanun's abrupt right turn. (A to C)** Khanun's track predicted by Pangu-Weather with (red) and without (blue) introducing initial perturbations in **(A)** both Region A and B (Exp-AB), **(B)** Region A only (Exp-A), and **(C)** Region B only (Exp-B). The black curve presents the observed Khanun's track. The domains of Region A and B are respectively denoted by rectangles in **(Fig. S4)**. **(D)** Predicted 500-hPa geopotential (contours, unit: gpm) in the unperturbed run (black) and Exp-AB (red), as well as their differences (Exp-AB minus unperturbed run, shadings, unit: gpm). **(E)** Same as **(D)**, except for 850-hPa wind field (vectors, unit: m/s) and differences in wind field (shadings, unit: m/s). **(F)** Same as **(D)**, except for the steering flow within 800 km radius from Khanun (vectors, unit: m/s) and differences in the zonal component of steering flow (shadings, unit: m/s). **(G)** Same as **(F)**, except for Exp-A. **(H to I)** Same as **(F)**, except for differences in **(H)** zonal and **(I)** meridional component of steering flow (shadings, unit: m/s) between Exp-AB and Exp-A (Exp-AB minus Exp-A). Panels **(D to L)** are derived by removing the circulation related to Khanun through vortex separation (see Methods).

## Similarities and differences between NWP and AI weather models

The above findings appear inconsistent with the findings of previous studies claiming that AI models failed to capture the butterfly effect (*31*, *32*). However, we note that they are not contradictory. In the following, we illustrate that two important criteria determining the appearance of butterfly effect in AI weather prediction: (1) the solution dependence on initial conditions, and (2) the magnitude of initial perturbations.

Firstly, in chaos theory, the sensitivity of forecast to initial perturbations appears only when initial states lie near the boundary between attractors (*48*). While the atmosphere is a well-known chaotic system, the butterfly effect is not observed at every single time step or for every initial condition (*49*). This characteristic is referred to as the solution dependence on initial conditions.

This criterion applies to both NWP and AI weather models. The state transition of both models' predictions appears at 1200 UTC 31 July (Figs. 1, S2), indicating that this specific initialization time lies near the boundary of the two local attractors. Predictions initialized at times away from this point are less sensitive to initial perturbations and exhibit smaller uncertainties (Fig. S2, S5) (*43*). For the same reason, the occurrence of the butterfly effect is also case-dependent. Only 14 out of 112 (~12%) TCs exhibit rapid adjustments in Pangu-Weather prediction (Fig. S1).

In contrast, the second criterion, i.e., the magnitude of initial perturbations, sets the difference in how the butterfly effect manifests between these two types of models. The spread of Pangu-Weather's ensemble output is dependent on the magnitude of initial perturbations. This dependency is evident by applying different evolution intervals ($\tau = i\Delta t$) in the fast physics-constrained perturbation generator, with a shorter $\tau$ corresponding to smaller initial perturbations (see Methods). Note that the magnitudes of initial perturbations $\tau < 24h$ are smaller than Pangu-Weather's model integration tendency.

When $\tau = 1h$, most ensemble members predict a right-turn trajectory, which is consistent with the control run. Only 1 member predicts a northwestward track of Khanun. Correspondingly, only a single probability center of the predicted TC position is obtained under these configurations (Fig. 4A). These indicate that most ensemble members follow the same attractor dynamics as the control run, implying that initial perturbations are too weak to trigger state transition of the predicted systems.

When $\tau \geq 3h$, the initial perturbations are strong enough to trigger the butterfly effect. For initial perturbation magnitude comparable to model integration tendency ($\tau \geq 24h$), a larger number of ensemble members following the alternate attractor dynamics that drive Khanun's northwestward movement. The number of ensemble members predicting Khanun's northwestward track and its subsequent landfall greatly rises to 126 (24%) and 171 (32%), respectively for $\tau = 24h$ and $36h$. Concurrently, the probability center of the predicted TC position splits into two (Figs. 4E–4F), indicating that a subset of ensemble members has transitioned to the alternate local attractor under this perturbation strength. Importantly, this chaotic behavior is also simulated when the perturbation magnitude is smaller than the model's integration tendency. For $\tau = 3h, 6h, 12h$, the number of ensemble members predicting a northwestward track of Khanun reach 86 (16%), 91 (17%), and 92 (18%), respectively. Similarly, two probability center patterns of the predicted TC position are clearly observed (Figs. 4B–4D).

The above results confirm that AI weather models capture chaotic dynamics similar to NWP models, except that stronger initial perturbations more effectively trigger the butterfly effect in AI models (*31*, *32*). More importantly, it is worth emphasizing that this does not imply the absence of butterfly effect in AI weather models because the perturbations required to induce such chaotic behavior are still weaker than the model integration tendency (see Discussion).

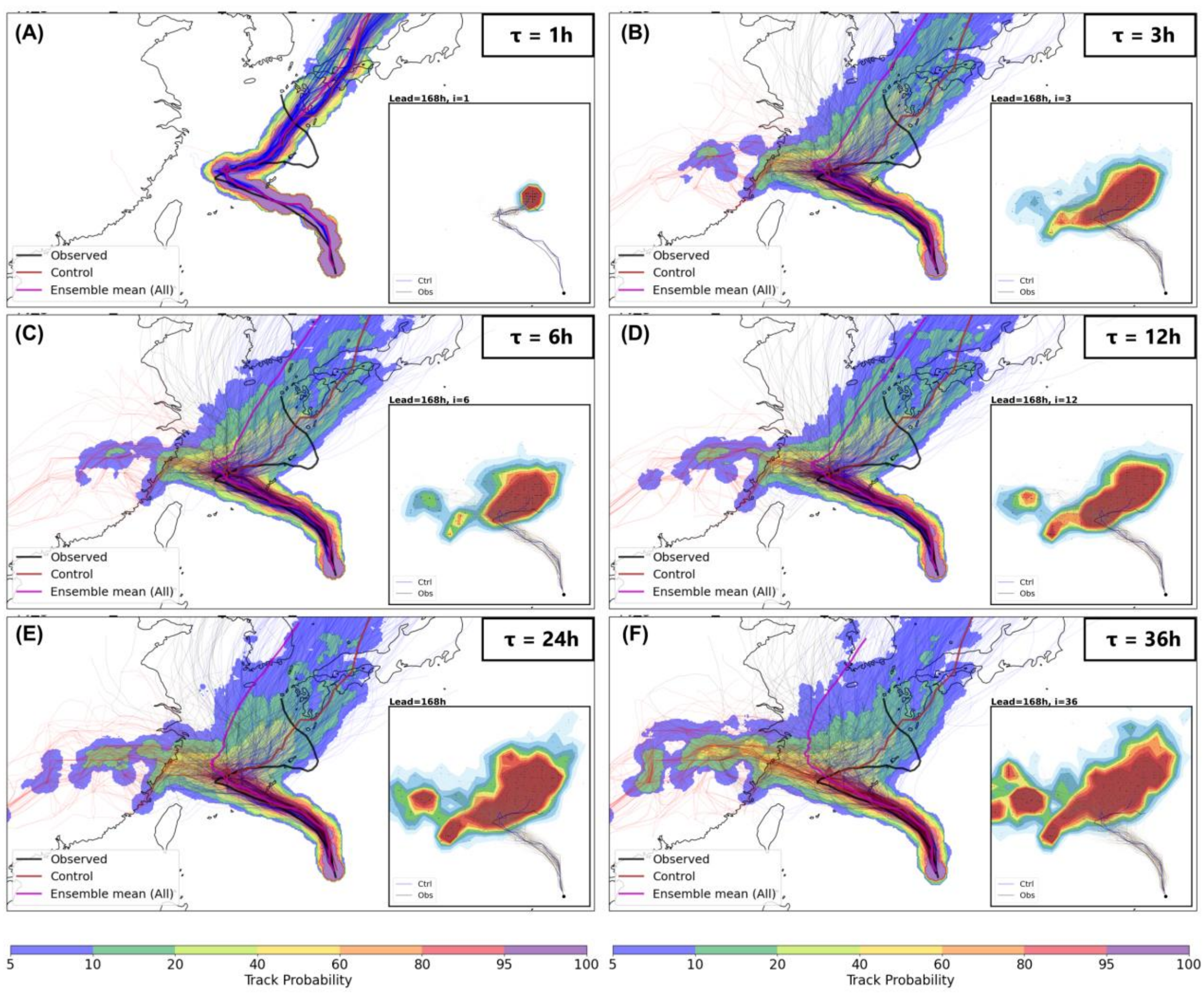

**Fig. 4. Sensitivity of Pangu-Weather's ensemble prediction spread to the magnitude of the introduced initial perturbations. (A to F)** Track probability (shading, unit: %) predicted by 528 ensemble members (thin curves) of Pangu-Weather on Khanun's abrupt right turn, with initialized at 1200UTC 30th July 2023. The black curve denotes the observed Khanun's trajectory. The red, blue, and purple curves respectively denote the averages of tracks moving northwestward, experiencing a right turn, and toward northward. The brown curve denotes the results of control run. The initial perturbations were generated by the fast physics-constrained perturbation generator (*27*), with the evolution interval of perturbation generator $\tau =$ **(A)** 1h, **(B)** 3h, **(C)** 6h, **(D)** 12h, **(E)** 24h, and **(F)** 36h, respectively. The small panels plot the probability distribution of predicted positions of Khanun at 168-hour forecast lead time.

## Discussion

This study demonstrates that state-of-the-art global AI weather models can indeed capture the butterfly effect. We show that Pangu-Weather's predictions on Super Typhoon Khanun are constrained by a double-attractor system. The perturbations required to trigger such chaotic behavior are weaker than the model integration tendency. Additionally, slight initial perturbations imposed over two key regions can substantially alter Khanun's predicted track, triggering state transitions between two local attractors.

AI weather models capture the butterfly effect through a mechanism distinct from NWP models. The primary reason for this is the different spatiotemporal resolution between AI and NWP models, particularly in their integration time steps (*50*). Current global AI weather models, designed with much longer time steps than NWP models, substantially reduce the total number of iterative computations to fewer than about 10 during the forecasting process. This inhibits the accumulation of error that enables rapid growth of small-scale perturbations in NWP simulations (*51–53*). Instead, AI models can capture butterfly effects triggered by perturbations with larger amplitudes and scales, which are more prevalent in real-world atmospheric systems (*54–56*).

In other words, the recent "paradox" originates from an unfair comparison between NWP and AI weather models. Global AI weather models are not designed to resolve small-scale perturbation evolution, compared to NWP models (*31*, *33*, *57*). This could be regarded as differences in representativeness errors between the two types of weather prediction models (*58–60*). It should be noted that such differences not only exist between AI and NWP models, but also exist among NWP models with different spatiotemporal resolutions (*31*). Thus, it is unfair to conclude the absence of butterfly effect in AI weather models solely due to the differences in representativeness errors between NWP and AI models.

More importantly, the rapid growth of small-scale perturbations is not the primary reason for the finite predictability of medium-range weather forecasts (*34*, *61*, *62*). Instead, uncertainty and chaoticity in short-to-medium-range forecasts are more closely linked to large-scale atmospheric dynamical processes (*34*, *54*). These suggest that the butterfly effect captured in AI models does not contradict that in NWP models.

This study verifies that AI weather models can simulate chaotic atmospheric behaviors, resolving the recent "paradox" in AI forecasting research. The butterfly effect determines the strong nonlinearity of atmospheric systems and is the fundamental problem of predicting weather evolution, especially those of extreme events such as TCs and extreme rainfall. Our results demonstrate that current AI models can effectively learn the chaotic behavior of large-scale atmospheric dynamics from training data, without any explicit physics assumptions. This highlights the great potential of data-driven AI methods for addressing challenges in weather forecasts and advancing the understanding of atmospheric dynamics. Furthermore, the existence of the butterfly effect in AI models validates the rationality and practical feasibility of AI-based ensemble prediction. The scale-dependent chaotic characteristics of AI models also provide practical guidance for future development of AI ensemble forecasts.

**References and Notes**

1. T. N. Palmer, Predicting uncertainty in forecasts of weather and climate. *Rep. Prog. Phys.* **63**, 71–116 (2000).

2. E. N. Lorenz, *The Essence of Chaos* (University of Washington press, Seattle, 1993)*The Jessie and John Danz lectures*.

3. S. Vannitsem, Predictability of large-scale atmospheric motions: Lyapunov exponents and error dynamics. *Chaos: An Interdisciplinary Journal of Nonlinear Science* **27**, 032101 (2017).

4. T. Gneiting, A. E. Raftery, Weather Forecasting with Ensemble Methods. *Science* **310**, 248–249 (2005).

5. T. Palmer, The real butterfly effect and maggoty apples. *Physics Today* **77**, 30–35 (2024).

6. P. Bauer, A. Thorpe, G. Brunet, The quiet revolution of numerical weather prediction. *Nature* **525**, 47–55 (2015).

7. F. Molteni, R. Buizza, T. N. Palmer, T. Petroliagis, The ECMWF Ensemble Prediction System: Methodology and validation. *Quart J Royal Meteoro Soc* **122**, 73–119 (1996).

8. R. Mureau, F. Molteni, T. N. Palmer, Ensemble prediction using dynamically conditioned perturbations. *Quart J Royal Meteoro Soc* **119**, 299–323 (1993).

9. Z. Toth, E. Kalnay, Ensemble Forecasting at NMC: The Generation of Perturbations. *Bull. Amer. Meteor. Soc.* **74**, 2317–2330 (1993).

10. M. S. Tracton, E. Kalnay, Operational Ensemble Prediction at the National Meteorological Center: Practical Aspects. *Wea. Forecasting* **8**, 379–398 (1993).

11. P. L. Houtekamer, L. Lefaivre, J. Derome, H. Ritchie, H. L. Mitchell, A System Simulation Approach to Ensemble Prediction. *Mon. Wea. Rev.* **124**, 1225–1242 (1996).

12. E. E. Ebert, Ability of a Poor Man's Ensemble to Predict the Probability and Distribution of Precipitation. *Mon. Wea. Rev.* **129**, 2461–2480 (2001).

13. Z. Li, D. Chen, The development and application of the operational ensemble prediction system at National Meteorological Center. *Journal of Applied Meteorological Science* **13**, 1–15 (2002).

14. F. A. Eckel, C. F. Mass, Aspects of Effective Mesoscale, Short-Range Ensemble Forecasting. *Weather and Forecasting* **20**, 328–350 (2005).

15. F.-C. Chien, Y.-C. Liu, B. J.-D. Jou, MM5 Ensemble Mean Forecasts in the Taiwan Area for the 2003 Mei-Yu Season. *Weather and Forecasting* **21**, 1006–1023 (2006).

16. W. J. Tennant, Z. Toth, K. J. Rae, Application of the NCEP Ensemble Prediction System to Medium-Range Forecasting in South Africa: New Products, Benefits, and Challenges. *Weather and Forecasting* **22**, 18–35 (2007).

17. J. Teixeira, C. A. Reynolds, K. Judd, Time Step Sensitivity of Nonlinear Atmospheric Models: Numerical Convergence, Truncation Error Growth, and Ensemble Design. *Journal of the Atmospheric Sciences* **64**, 175–189 (2007).

18. M. Matsueda, M. Kyouda, H. L. Tanaka, T. Tsuyuki, Daily Forecast Skill of Multi-Center Grand Ensemble. *SOLA* **3**, 29–32 (2007).

19. K. Bi, L. Xie, H. Zhang, X. Chen, X. Gu, Q. Tian, Accurate medium-range global weather forecasting with 3D neural networks. *Nature* **619**, 533–538 (2023).

20. R. Lam, A. Sanchez-Gonzalez, M. Willson, P. Wirnsberger, M. Fortunato, F. Alet, S. Ravuri, T. Ewalds, Z. Eaton-Rosen, W. Hu, A. Merose, S. Hoyer, G. Holland, O. Vinyals, J. Stott, A. Pritzel, S. Mohamed, P. Battaglia, Learning skillful medium-range global weather forecasting. *Science* **382**, 1416–1421 (2023).

21. L. Chen, X. Zhong, F. Zhang, Y. Cheng, Y. Xu, Y. Qi, H. Li, FuXi: a cascade machine learning forecasting system for 15-day global weather forecast. *npj Clim Atmos Sci* **6**, 190 (2023).

22. K. Chen, T. Han, J. Gong, L. Bai, F. Ling, J.-J. Luo, X. Chen, L. Ma, T. Zhang, R. Su, Y. Ci, B. Li, X. Yang, W. Ouyang, FengWu: Pushing the Skillful Global Medium-range Weather Forecast beyond 10 Days Lead. arXiv [Preprint] (2023). https://doi.org/10.48550/ARXIV.2304.02948.

23. J. Pathak, S. Subramanian, P. Harrington, S. Raja, A. Chattopadhyay, M. Mardani, T. Kurth, D. Hall, Z. Li, K. Azizzadenesheli, P. Hassanzadeh, K. Kashinath, A. Anandkumar, FourCastNet: A Global Data-driven High-resolution Weather Model using Adaptive Fourier Neural Operators. arXiv [Preprint] (2022). https://doi.org/10.48550/ARXIV.2202.11214.

24. I. Price, A. Sanchez-Gonzalez, F. Alet, T. R. Andersson, A. El-Kadi, D. Masters, T. Ewalds, J. Stott, S. Mohamed, P. Battaglia, R. Lam, M. Willson, Probabilistic weather forecasting with machine learning. *Nature* **637**, 84–90 (2025).

25. S. Lang, M. Alexe, M. C. A. Clare, C. Roberts, R. Adewoyin, Z. B. Bouallègue, M. Chantry, J. Dramsch, P. D. Dueben, S. Hahner, P. Maciel, A. Prieto-Nemesio, C. O'Brien, F. Pinault, J. Polster, B. Raoult, S. Tietsche, M. Leutbecher, AIFS-CRPS: Ensemble forecasting using a model trained with a loss function based on the Continuous Ranked Probability Score. arXiv [Preprint] (2024). https://doi.org/10.48550/ARXIV.2412.15832.

26. X. Zhong, L. Chen, H. Li, J. Liu, X. Fan, J. Feng, K. Dai, J.-J. Luo, J. Wu, B. Lu, FuXi-ENS: A machine learning model for medium-range ensemble weather forecasting. arXiv [Preprint] (2024). https://doi.org/10.48550/ARXIV.2405.05925.

27. J. Pu, M. Mu, J. Feng, X. Zhong, H. Li, A fast physics-based perturbation generator of machine learning weather model for efficient ensemble forecasts of tropical cyclone track. *npj Clim Atmos Sci* **8**, 128 (2025).

28. J. Baño-Medina, A. Sengupta, D. Watson-Parris, W. Hu, L. Delle Monache, Toward Calibrated Ensembles of Neural Weather Model Forecasts. *J Adv Model Earth Syst* **17**, e2024MS004734 (2025).

29. J. Du, S. SADEGHI TABAS, J. Wang, J. Carley, Understanding similarities and differences between data-driven AI model-based and physics model-based ensemble forecasts. *NCEP Office Notes*, 31 (2025).

30. D. M. McAfee, E. A. Barnes, Am I confused or is this confusing?: Deep ensembles for ENSO uncertainty quantification. *Mach. Learn.: Earth* **2**, 015012 (2026).

31. T. Selz, G. C. Craig, Can Artificial Intelligence-Based Weather Prediction Models Simulate the Butterfly Effect? *Geophysical Research Letters* **50**, e2023GL105747 (2023).

32. H. Kim, J. Ryu, S.-W. Son, J.-H. Jeong, H. Kim, J.-H. Yoon, A spectral test of the butterfly effect and physical consistency in the diffusion-based GenCast's ensembles. *npj Clim Atmos Sci*, doi: 10.1038/s41612-026-01380-1 (2026).

33. M. Mu, B. Qin, G. Dai, Predictability Study of Weather and Climate Events Related to Artificial Intelligence Models. *Adv. Atmos. Sci.* **42**, 1–8 (2025).

34. B.-W. Shen, R. Pielke Sr., X. Zeng, Butterfly Effects and Finite Predictability in AI-Based Weather Prediction. *ESS OPEN ARCHIVE* (2025).

35. H. Hersbach, B. Bell, P. Berrisford, S. Hirahara, A. Horányi, J. Muñoz-Sabater, J. Nicolas, C. Peubey, R. Radu, D. Schepers, A. Simmons, C. Soci, S. Abdalla, X. Abellan, G. Balsamo, P. Bechtold, G. Biavati, J. Bidlot, M. Bonavita, G. De Chiara, P. Dahlgren, D. Dee, M. Diamantakis, R. Dragani, J. Flemming, R. Forbes, M. Fuentes, A. Geer, L. Haimberger, S. Healy, R. J. Hogan, E. Hólm, M. Janisková, S. Keeley, P. Laloyaux, P. Lopez, C. Lupu, G. Radnoti, P. De Rosnay, I. Rozum, F. Vamborg, S. Villaume, J. Thépaut, The ERA5 global reanalysis. *Quart J Royal Meteoro Soc* **146**, 1999–2049 (2020).

36. E. N. Lorenz, The predictability of a flow which possesses many scales of motion. *Tellus* **21**, 289–307 (1969).

37. J. Pathak, Z. Lu, B. R. Hunt, M. Girvan, E. Ott, Using machine learning to replicate chaotic attractors and calculate Lyapunov exponents from data. *Chaos: An Interdisciplinary Journal of Nonlinear Science* **27**, 121102 (2017).

38. C. Kieu, Searching for Chaos in Tropical Cyclone Intensity: A Machine Learning Approach. *Tellus A: Dynamic Meteorology and Oceanography* **76**, 166–176 (2024).

39. R. Godwin-Jones, Chasing the butterfly effect: Informal language learning online as a complex system. *LLT* **22**, 8–27 (2018).

40. H. Sabzian, N. Shahriari, M. G. Nejad, Unraveling the Butterfly Effects in Social Dynamics: Insights from Agent-Based Modeling. arXiv arXiv:2312.07914 [Preprint] (2023). https://doi.org/10.48550/arXiv.2312.07914.

41. Japan Meteorological Agency, "Annual Report on the Activities of the RSMC Tokyo - Typhoon Center 2023" (Japan Meteorological Agency).

42. D. Xu, Z. Lu, J. C.-H. Leung, D. Zhao, Y. Li, Y. Shi, B. Chen, G. Nie, N. Wu, X. Tian, Y. Yang, S. Zhang, B. Zhang, AI models still lag behind traditional numerical models in predicting sudden-turning typhoons. *Science Bulletin* **70**, 2705–2708 (2025).

43. Yiwu Huang, Lin Dong, Qifeng Qian, Xinyan Lu, Analysis on challenges and improvement strategies in track forecasting of Typhoon Khanun (2306). *Journal of Marine Meteorology* **1**, 64–74 (2026).

44. Y. Shi, R. Hu, N. Wu, H. Zhang, X. Liu, Z. Zeng, J. Zhu, P. Han, C. Luo, H. Zhang, J. He, X. Shi, Comparison of AI and NWP Models in Operational Severe Weather Forecasting: A Study on Tropical Cyclone Predictions. *Journal of Geophysical Research: Machine Learning and Computation* **2**, e2024JH000481 (2025).

45. V. I. Oseledec, A multiplicative ergodic theorem: Lyapunov characteristic num-bers for dynamical systems. *Trans Moscow Math Soc* **19** (1968).

46. B. Chen, J. Li, R. Ding, Nonlinear local Lyapunov exponent and atmospheric predictability research. *SCI CHINA SER D* **49**, 1111–1120 (2006).

47. R.-Q. Ding, J.-P. Li, K.-J. Ha, Nonlinear Local Lyapunov Exponent and Quantification of Local Predictability. *Chinese Phys. Lett.* **25**, 1919–1922 (2008).

48. C. Grebogi, S. W. McDonald, E. Ott, J. A. Yorke, Final state sensitivity: An obstruction to predictability. *Physics Letters A* **99**, 415–418 (1983).

49. B.-W. Shen, R. A. Pielke, X. Zeng, J.-J. Baik, S. Faghih-Naini, J. Cui, R. Atlas, T. A. L. Reyes, "Is Weather Chaotic? Coexisting Chaotic and Non-chaotic Attractors Within Lorenz Models" in *13th Chaotic Modeling and Simulation International Conference*, C. H. Skiadas, Y. Dimotikalis, Eds. (Springer International Publishing, Cham, 2021; https://link.springer.com/10.1007/978-3-030-70795-8_57)*Springer Proceedings in Complexity*, pp. 805–825.

50. C. Kieu, Predictability of Global AI Weather Models. arXiv arXiv:2410.03266 [Preprint] (2024). https://doi.org/10.48550/arXiv.2410.03266.

51. ECMWF, IFS Documentation CY41R2 - Part III: Dynamics and Numerical Procedures. doi: 10.21957/83WOUV80 (2016).

52. L. Wang, Y. Liu, D. Xu, L. Zhang, J. C. Leung, H. Li, J. Gong, B. Zhang, An INCREMENTAL ANALYSIS UPDATE IN THE framework OF THE FOUR-DIMENSIONAL VARIATIONAL DATA ASSIMILATION : Description and preliminary tests in the operational China Meteorological Administration Global Forecast System. *Quart J Royal Meteoro Soc* **150**, 2104–2122 (2024).

53. X. Zhou, Y. Zhu, D. Hou, Y. Luo, J. Peng, R. Wobus, Performance of the New NCEP Global Ensemble Forecast System in a Parallel Experiment. *Weather and Forecasting* **32**, 1989–2004 (2017).

54. D. R. Durran, M. Gingrich, Atmospheric Predictability: Why Butterflies Are Not of Practical Importance. *Journal of the Atmospheric Sciences* **71**, 2476–2488 (2014).

55. N. Bei, F. Zhang, Impacts of initial condition errors on mesoscale predictability of heavy precipitation along the Mei-Yu front of China. *Quart J Royal Meteoro Soc* **133**, 83–99 (2007).

56. D. R. Durran, P. A. Reinecke, J. D. Doyle, Large-Scale Errors and Mesoscale Predictability in Pacific Northwest Snowstorms. *Journal of the Atmospheric Sciences* **70**, 1470–1487 (2013).

57. M. Bonavita, On Some Limitations of Current Machine Learning Weather Prediction Models. *Geophysical Research Letters* **51**, e2023GL107377 (2024).

58. T. Janjić, N. Bormann, M. Bocquet, J. A. Carton, S. E. Cohn, S. L. Dance, S. N. Losa, N. K. Nichols, R. Potthast, J. A. Waller, P. Weston, On the representation error in data assimilation. *Quart J Royal Meteoro Soc* **144**, 1257–1278 (2018).

59. B. Tustison, D. Harris, E. Foufoula-Georgiou, Scale issues in verification of precipitation forecasts. *J. Geophys. Res.* **106**, 11775–11784 (2001).

60. M. Göber, E. Zsótér, D. S. Richardson, Could a perfect model ever satisfy a naïve forecaster? On grid box mean *versus* point verification. *Meteorological Applications* **15**, 359–365 (2008).

61. T. Selz, M. Riemer, G. C. Craig, The Transition from Practical to Intrinsic Predictability of Midlatitude Weather. *Journal of the Atmospheric Sciences* **79**, 2013–2030 (2022).

62. T. N. Palmer, A. Döring, G. Seregin, The real butterfly effect. *Nonlinearity* **27**, R123–R141 (2014).

63. M. Ying, W. Zhang, H. Yu, X. Lu, J. Feng, Y. Fan, Y. Zhu, D. Chen, An Overview of the China Meteorological Administration Tropical Cyclone Database. *Journal of Atmospheric and Oceanic Technology* **31**, 287–301 (2014).

64. E. N. Lorenz, Atmospheric predictability experiments with a large numerical model. *Tellus A: Dynamic Meteorology and Oceanography* **34**, 505 (1982).

65. D. Xu, J. C.-H. Leung, B. Zhang, A Time Neighborhood Method for the Verification of Landfalling Typhoon Track Forecast. *Adv. Atmos. Sci.* **40**, 273–284 (2023).

66. H. Van Nguyen, Y.-L. Chen, High-Resolution Initialization and Simulations of Typhoon Morakot (2009). *Mon. Wea. Rev.* **139**, 1463–1491 (2011).

**Acknowledgments:** The authors would like to thank Prof. Weihong Qian from Peking University for their valuable comments. We thank the technical support of the National Large Scientific and Technological Infrastructure "Earth System Numerical Simulation Facility" (https://cstr.cn/31134.02.EL).

**Funding:**

Innovation Research Foundation of National University of Defense Technology (202402-YJRC-LJ-001)

National Natural Science Foundation of China (42405038)

**Author contributions:**

Conceptualization: BZ, QZ

Methodology: JCHL, DX, BZ

Formal analysis: JCHL, DX, WY, SZ, XZ, YL, BZ

Visualization: JCHL, DX

Data Curation: GN, WY, JCHL

Software: JF, JP, WY, GN, SZ, JCHL

Funding acquisition: BZ, YL, KR, JCHL

Supervision: BZ, YL, QZ

Administration: YL, KR

Writing – original draft: JCHL, BZ, DX

Writing – review & editing: JCHL, DX, XZ, WY, SZ, GN, JF, JP, YL, KR, QZ, BZ

**Competing interests:** The authors declare no competing interests.

**Data and materials availability:** The CMA Tropical Cyclone Best Track dataset is accessible through https://tcdata.typhoon.org.cn/en/zjljsjj.html. The ERA5 is accessible through https://cds.climate.copernicus.eu/datasets/reanalysis-era5-pressure-levels?tab=download. The Pangu-Weather model is accessible through https://github.com/198808xc/Pangu-Weather.

**Supplementary Materials**

Materials and Methods

Supplementary Text S1 to S2

Figs. S1 to S5

References (62–65)

# Supplementary Information

**The PDF file includes:**

Materials and Methods

Supplementary Text S1 to S2

Figs. S1 to S5

References (62–65)

## Materials and Methods

### Data

Two datasets were used in the following analysis. The European Centre for Medium-Range Weather Forecasts (ECMWF) Reanalysis v5 (ERA5) (*35*) provided initial conditions for global AI weather models. The ERA5 is the training dataset for most current global AI weather models. The (CMA) Tropical Cyclone Best Track dataset (*63*) provided TC observation data, including trajectories, which was used as ground truth to evaluate global AI weather model forecast errors.

### AI Weather Model

In this study, we employ the Pangu-Weather model to explore the existence of the butterfly effect within it. Pangu-Weather was among the first published global AI weather models and exhibits outstanding forecast skills for global-scale circulation, with particularly good performance in TC trajectory prediction (*19*, *42*). Pangu-Weather employs a hierarchical temporal aggregation approach to reduce cumulative forecast errors, by calling the base deep networks (lead times being 1 h, 3 h, 6 h or 24 h) iteratively, using each forecasted result as the input of the next step. Take a 36-hour prediction as an example, the 24-hour forecast model is executed once, and the 6-hour forecast model is executed twice based on the 24-hour forecast output.

It is also the model analyzed in Selz and Craig (2023), who drew the conclusion that AI weather models cannot reproduce the atmosphere's butterfly effect. In this study, ERA5 reanalysis was used as the initial condition in all experiments based on the Pangu-Weather model.

### AI Ensemble Forecast

The ensemble approach was applied to Pangu-Weather to test the sensitivity of its prediction to initial perturbation. We employed the fast physics-constrained perturbation generator (*27*) to generate initial ensemble perturbations for probabilistic AI weather prediction experiments. The core idea of this approach is to apply perturbations with appropriate amplitudes and physical constraints, so that the growth rate of perturbation evolution in AI weather models is comparable to that in NWP models.

Technically, preliminary physical-constraint perturbations were first sampled based on the lagged forecast method (*64*). Suppose we have two forecasts: (1) $F_{t_{-i},(i+j)\Delta t}$ initialized at time $t_{-i}$ forecast leading time of $(i+j)\Delta t$, and (2) $F_{t_0,j\Delta t}$ initialized at time $t_0$ and forecast leading time of $j\Delta t$, where $\Delta t$ represents the time interval. We define the difference between the two forecasts as $e(i\Delta t, j\Delta t)$. For each initial time $t_0$ of prediction, we sampled the model's own successively evolved initial perturbations within a within a short-time window from $t_0 - n\Delta t$ to $t_0$, denoted as $(e(i\Delta t, 0)|_{t_0}, e(i\Delta t, 0)|_{t_{-1}}, \dots, e(i\Delta t, 0)|_{t_{-n}})$, where the subscript $t_{-n}$ indicates that the evolved perturbation is valid at time $t_0 - n\Delta t$. In this study, perturbations $e(i\Delta t, 0)$ were calculated at 12-hour intervals ($\Delta t = 12h$), with the time window set to 12 days before the initial time $t_0$. The evolution interval ($\tau = i\Delta t$) was tested in a range of values, including 1 h, 3 h, 6 h, 12 h, 24 h, and 36 h. These preliminary initial perturbations $e(i\Delta t, 0)$ were paired as positive-negative pairs to generate initial ensemble perturbations. These perturbations were then added to the ERA5 reanalysis field to create perturbed initial states, or the input of ensemble AI weather forecasts. In the presented analysis, each probabilistic prediction experiment involves 528 ensemble members, in addition to the control run. $\tau = 24$ is applied if without explicitly specified. For more details about the fast physics-constrained perturbation generator, readers are referred to Pu et al. (2025).

Evaluation of TC Track Prediction

To compare the differences between two forecasts or between forecasts and observations, we adopted the traditional point-to-point TC track forecast evaluation method. This is done by simply computing the great circle distance between positions of two storms, defined by longitude and latitude, for each forecast lead time. While research has noted that this traditional approach may erroneously include information about differences in cyclone moving speed (*65*), our analyses indicate that this does not affect the key conclusion of this study.

Nonlinear Local Lyapunov Exponent

In weather forecast applications, a simple yet widely adopted method to quantify chaos is examining the growth of prediction errors and their spread. Besides, the Lyapunov exponent is an effective indicator to quantify the existence of chaos, or butterfly effect, in a system. In this study, we employ the so-called nonlinear local Lyapunov exponent (NLLE), which is designed to measures the growth rate of initial errors of nonlinear dynamical models without linearizing the governing equations (*46*, *47*). The NLLE ($\lambda$) of a forecast at initial time $t_0$ is defined as

$$\lambda(\boldsymbol{x}(t_0), \boldsymbol{\delta}(t_0), \tau) = \frac{1}{\tau} \ln \frac{\|\boldsymbol{\delta}(t_0+\tau)\|}{\|\boldsymbol{\delta}(t_0)\|} \quad (1)$$

where $\boldsymbol{x}(t_0)$ denotes the initial state, $\boldsymbol{\delta}(t_0)$ is the initial perturbation, $\boldsymbol{\delta}(t_0 + \tau)$ is the perturbation at forecast lead time $\tau$. The local ensemble mean of the NLLE can be derived by averaging $\lambda$ in Eq. (1) over a large number of ensemble members.

Vortex separation

In order to investigate the environmental background circulation that influences TC track, a vortex separation approach (*66*) was applied to remove TC-related signals from the atmospheric fields, such as geopotential and wind fields.

Technically, a variable $F$ is decomposed into two components, the vortex component $F^{vor}$ and the environment component $F^{env}$ (Eq. 2).

$$F = F^{env} + F^{vor} \quad (2)$$

The $i^{th}$ and $j^{th}$ denotes grid point of environment component $F^{env}$ is derived by the following formulation:

$$\overline{F_{i,j}^{env}} = F_{i,j} + K_m\left(F_{i,j-q_n} - 2F_{i,j} + F_{i,j+q_n}\right) \quad (3)$$

$$F_{i,j}^{env} = \overline{F_{i,j}^{env}} + K_m\left(\overline{F_{i,j-q_n}^{env}} - 2\overline{F_{i,j}^{env}} + \overline{F_{i,j+q_n}^{env}}\right) \quad (4)$$

$$q_n = \left[\frac{111\cos(\phi_0)}{n\Delta}\right] \quad n = 1,2,4, \dots, M \quad (5)$$

$$M = \left[\frac{111\cos(\phi_0)}{\Delta}\right] \quad (6)$$

$$K_m = \left[\frac{1}{2}\left(1 - \cos\frac{2\pi}{m}\right)\right] \quad m = 2,3,4,2,5,6,7,2,8,9,2 \quad (7)$$

where $\phi_0$ denotes the TC center latitude (unit: rad), $\Delta$ denotes the horizonal resolution of atmospheric data (unit: km), [ ] represents the nearest integer number. Analyses presented in Fig. 3 are the derived outputs of $F^{env}$.

Composite analysis and significance test

The statistical significance (p values) of all composite analyses presented in this paper was tested by the Student's *t*-test. All composite values presented in the main context are statistically significant at a confidence level of 99.99% (p value $\leq$ 0.0001), unless otherwise specified.

## Supplementary Information

Text S1. Super Typhoon Khanun (2023)

Khanun was distinguished by its unusual trajectory, which is characterized by its two sharp turns within 5 days when it passed through the Ryukyu Islands. The typhoon first formed southwest of Guam on 22$^{nd}$ July 2023, then moved northwestward toward China under the control of western Pacific subtropical high (WPSH). Dramatically, while all operational weather forecast models and agencies predicted Khanun would make landfall along the east coast of China, the typhoon suddenly slowed down and made an abrupt right turn at 1200UTC 3$^{rd}$ August. On 7$^{th}$ August, Khanun experienced its second sharp turn, resuming a northwestward track, and resulting in a zigzagging path. The typhoon finally made landfall in Korea on 10$^{th}$ August, 19 days after its genesis (Fig. 1A). Due to its extremely unusual trajectory, combined with strong intensity and a long lifetime, Khanun caused substantial damages in surrounding regions, including Japan, China, North Korea, and South Korea, etc. (*41*).

Text S2. WNP TC cases that exhibit rapid adjustments in the predicted TC tracks

We performed hindcast experiment on 112 WNP TC cases based on Pangu-Weather. The results show that 14 out of 112 (~12%) cases exhibit rapid adjustments in the predicted TC tracks within a short interval of initialization times. These 14 cases are respectively Vongfong and Chan-hom in 2020; Surigae, Choi-wan, Chanthu, and Namtheun in 2021; Muifa in 2022; Khanun, Haikui, and Kirogi in 2023; Pulasan, Krathon, Kong-rey, and Usagi in 2023 (Fig. S1).

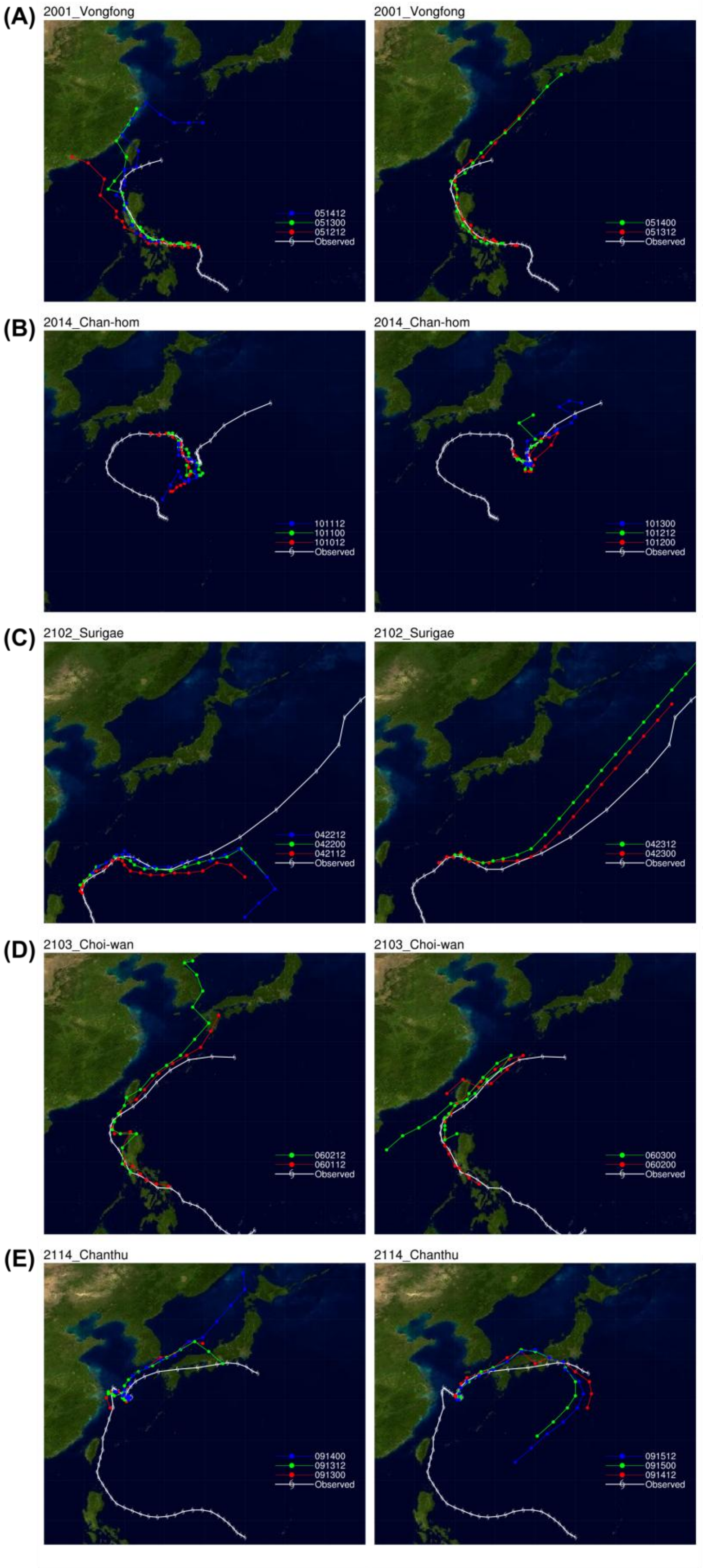
(A)
2001_Vongfong
051412
051300
051212
Observed
2001_Vongfong
051400
051312
Observed
(B)
2014_Chan-hom
101112
101100
101012
Observed
2014_Chan-hom
101300
101212
101200
Observed
(C)
2102_Surigae
042212
042200
042112
Observed
2102_Surigae
042312
042300
Observed
(D)
2103_Choi-wan
060212
060112
Observed
2103_Choi-wan
060300
060200
Observed
(E)
2114_Chanthu
091400
091312
091300
Observed
2114_Chanthu
091512
091500
091412
Observed

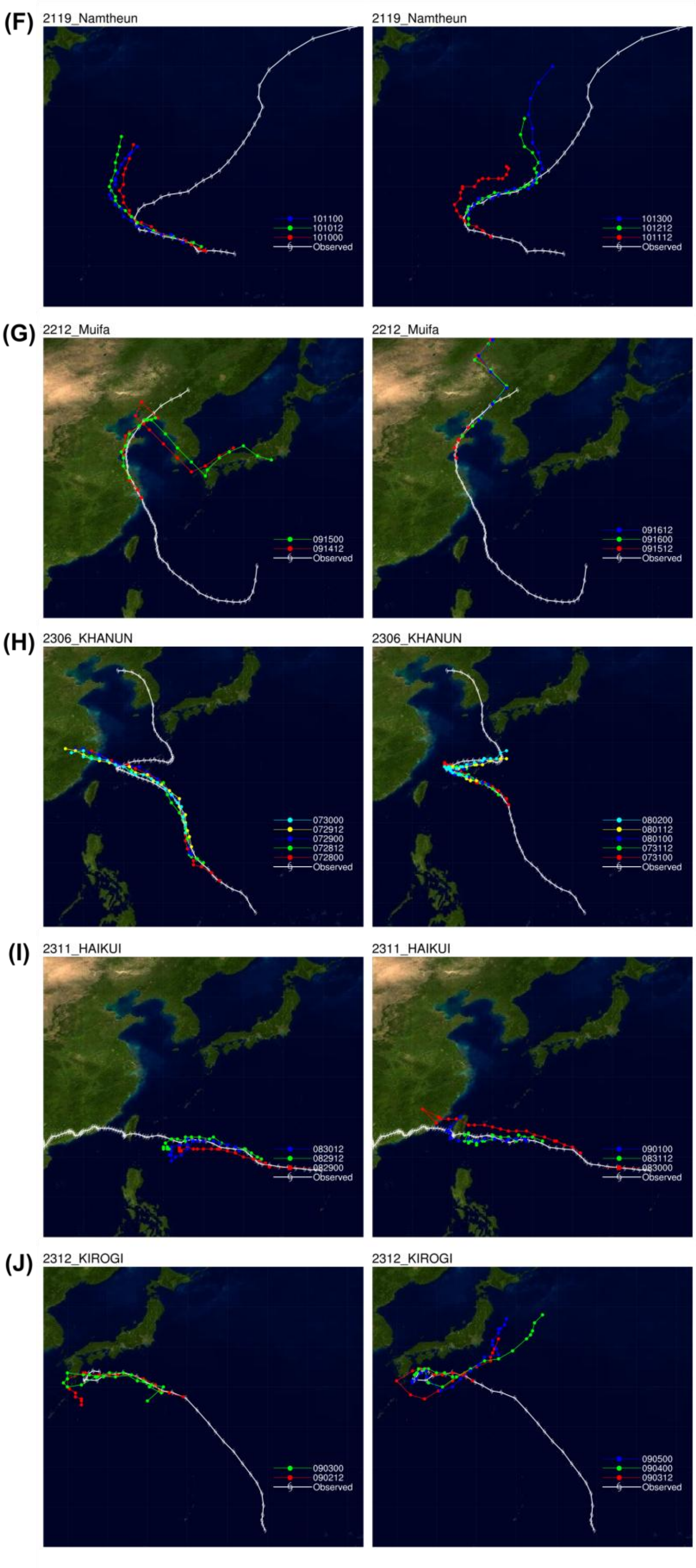
(F)
2119_Namtheun
101100
101012
101000
Observed
2119_Namtheun
101300
101212
101112
Observed
(G)
2212_Muifa
091500
091412
Observed
2212_Muifa
091612
091600
091512
Observed
(H)
2306_KHANUN
073000
072912
072900
072812
072800
Observed
2306_KHANUN
080200
080112
080100
073112
073100
Observed
(I)
2311_HAIKUI
083012
082912
082900
Observed
2311_HAIKUI
090100
083112
083000
Observed
(J)
2312_KIROGI
090300
090212
Observed
2312_KIROGI
090500
090400
090312
Observed

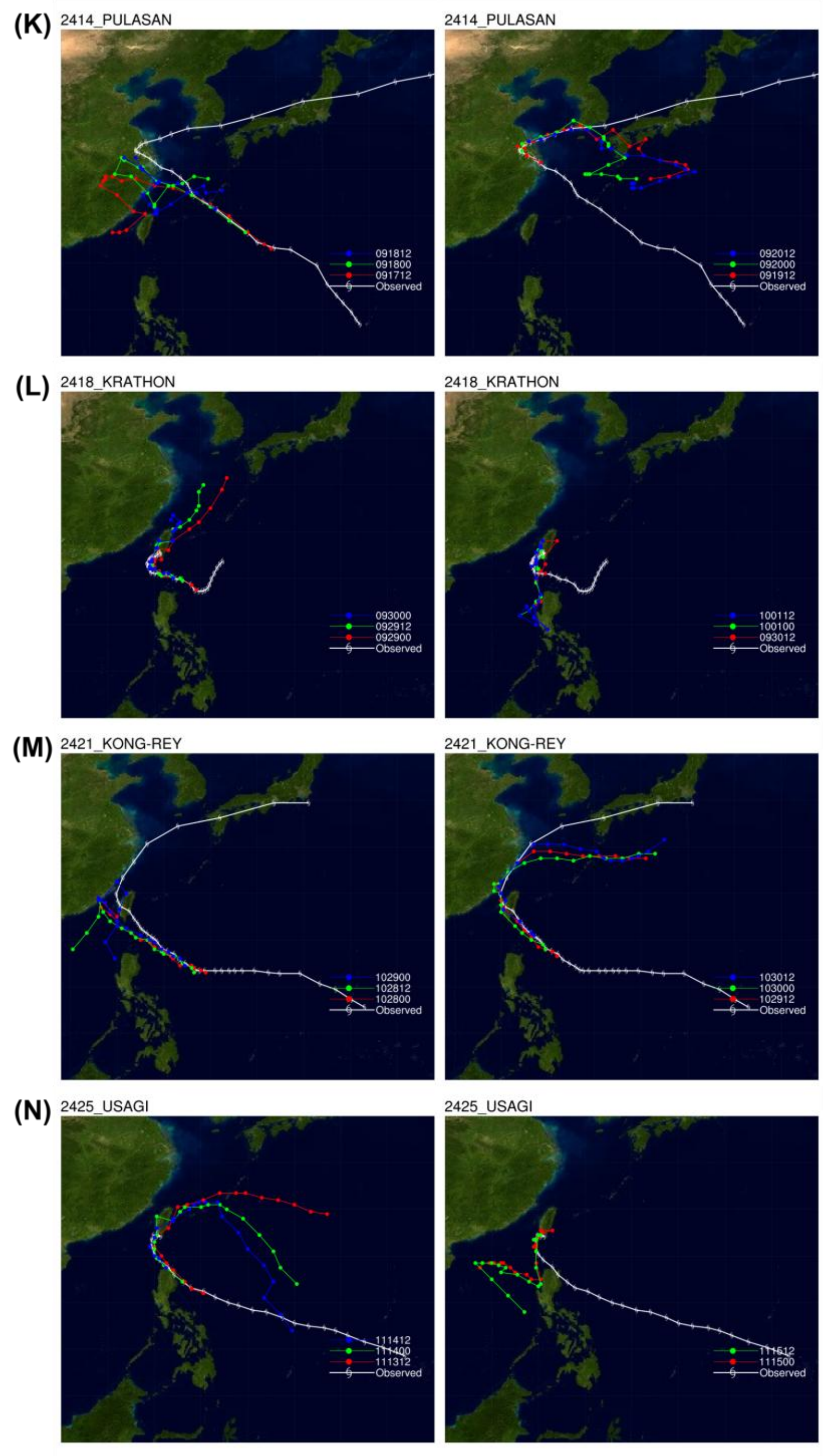

**Fig. S1. WNP TCs that exhibit rapid adjustments in the predicted tracks.** Pangu-Weather predicted tracks of TC **(A)** Vongfong (2020), **(B)** Chan-hom (2020), **(C)** Surigae (2021), **(D)** Choi-wan (2021), **(E)** Chanthu (2021), **(F)** Namtheun (2021), **(G)** Muifa (2022); **(H)** Khanun (2023), **(I)** Haikui (2023), **(J)** Kirogi (2023), **(K)** Pulasan (2024), **(L)** Krathon (2024), **(M)** Kong-rey (2024), and **(N)** Usagi (2024), respectively, based on Pangu-Weather at different initialization times.

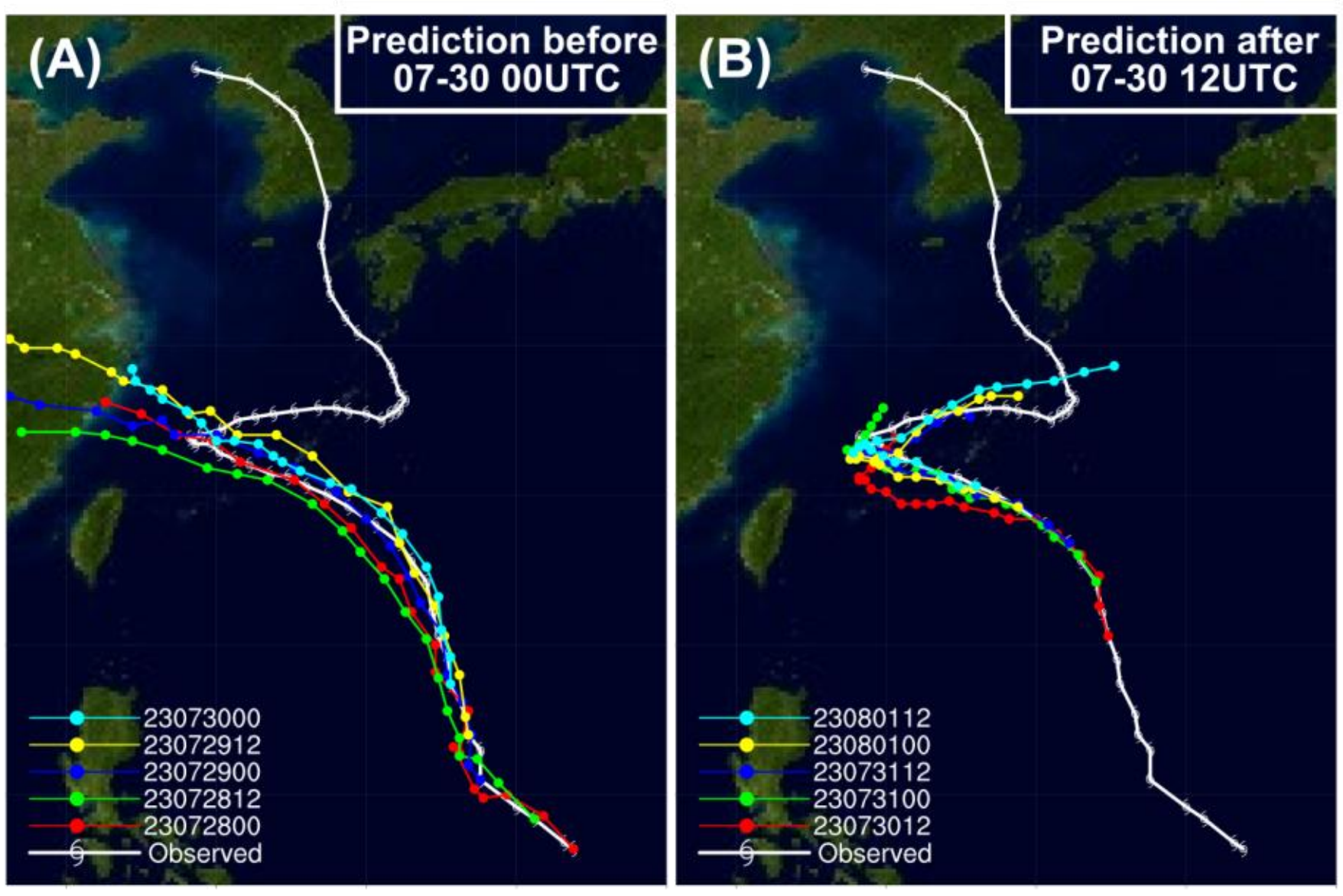

**Fig. S2. Rapid adjustment of NWP simulation results in the abrupt right turn of Super Typhoon Khanun.** Same as Figs. 1B–1C in the main text, except for forecast results of ECMWF-IFS.

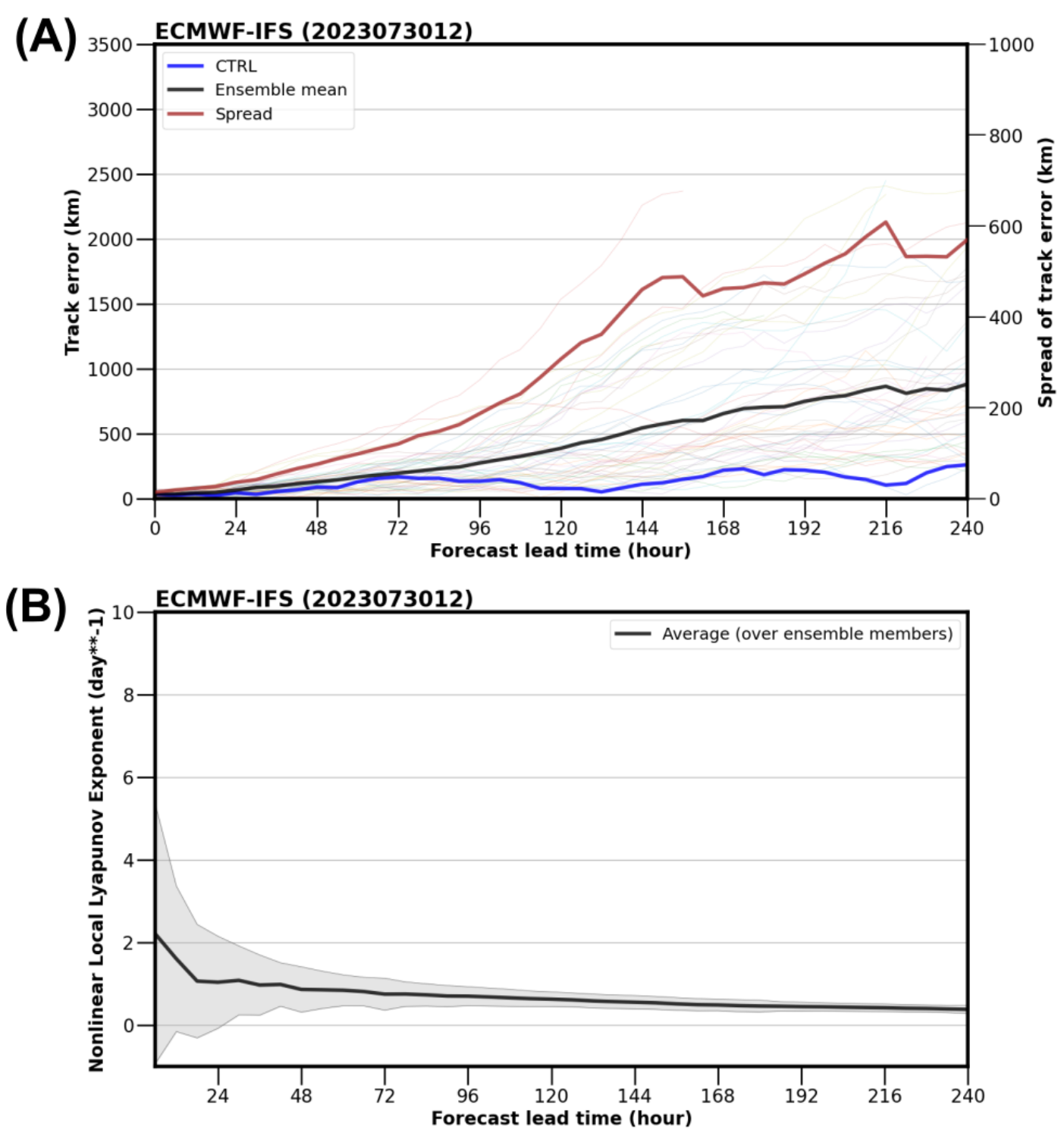

**Fig. S3. Spread of NWP probabilistic prediction on Khanun's abrupt right turn.** Same as Figs. 2E and 2F in the main text, except for ensemble results of ECMWF-IFS.

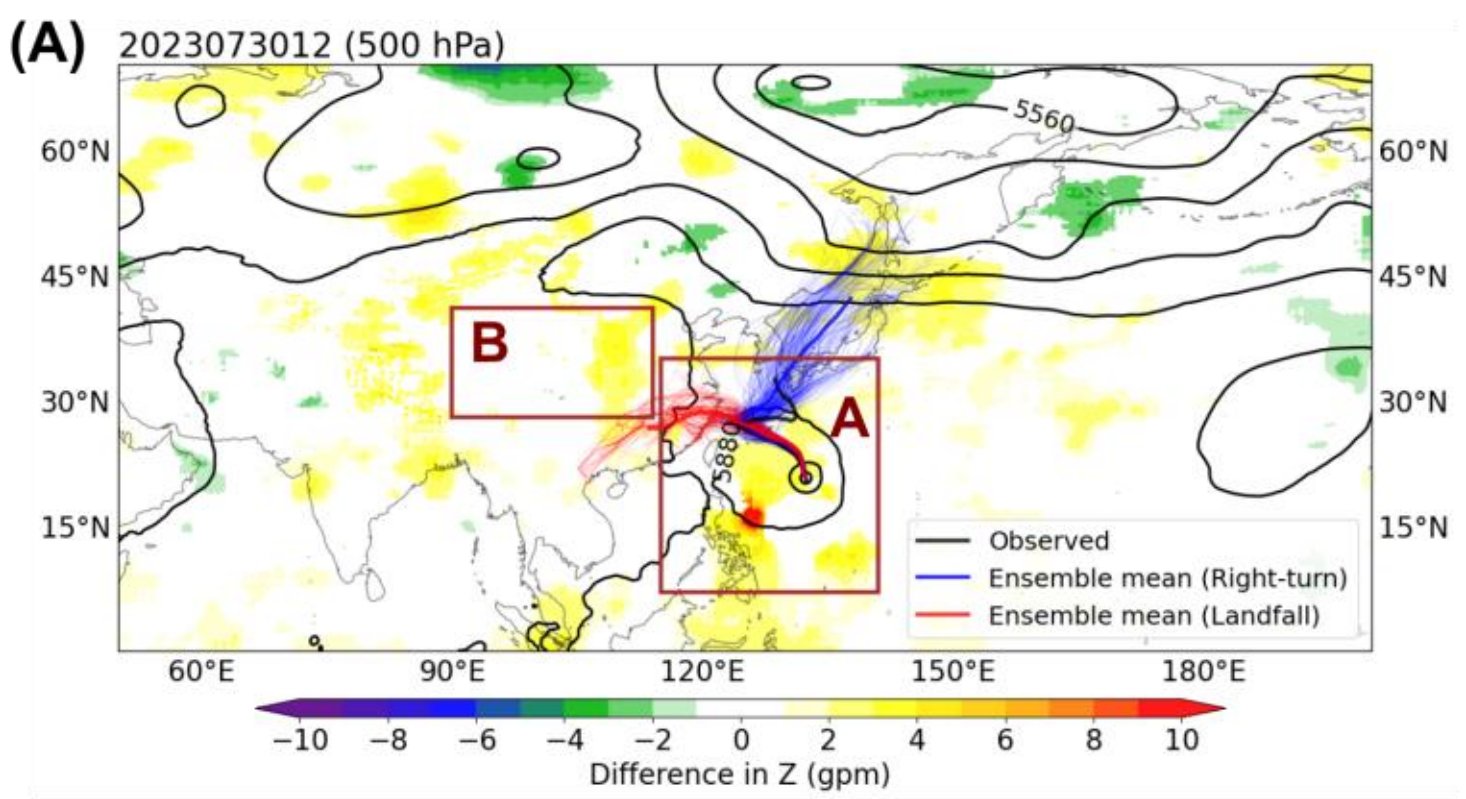
(A)
2023073012 (500 hPa)
5560
5880
B
A
Observed
Ensemble mean (Right-turn)
Ensemble mean (Landfall)
60°N
45°N
30°N
15°N
60°E
90°E
120°E
150°E
180°E
−10 −8 −6 −4 −2 0 2 4 6 8 10
Difference in Z (gpm)

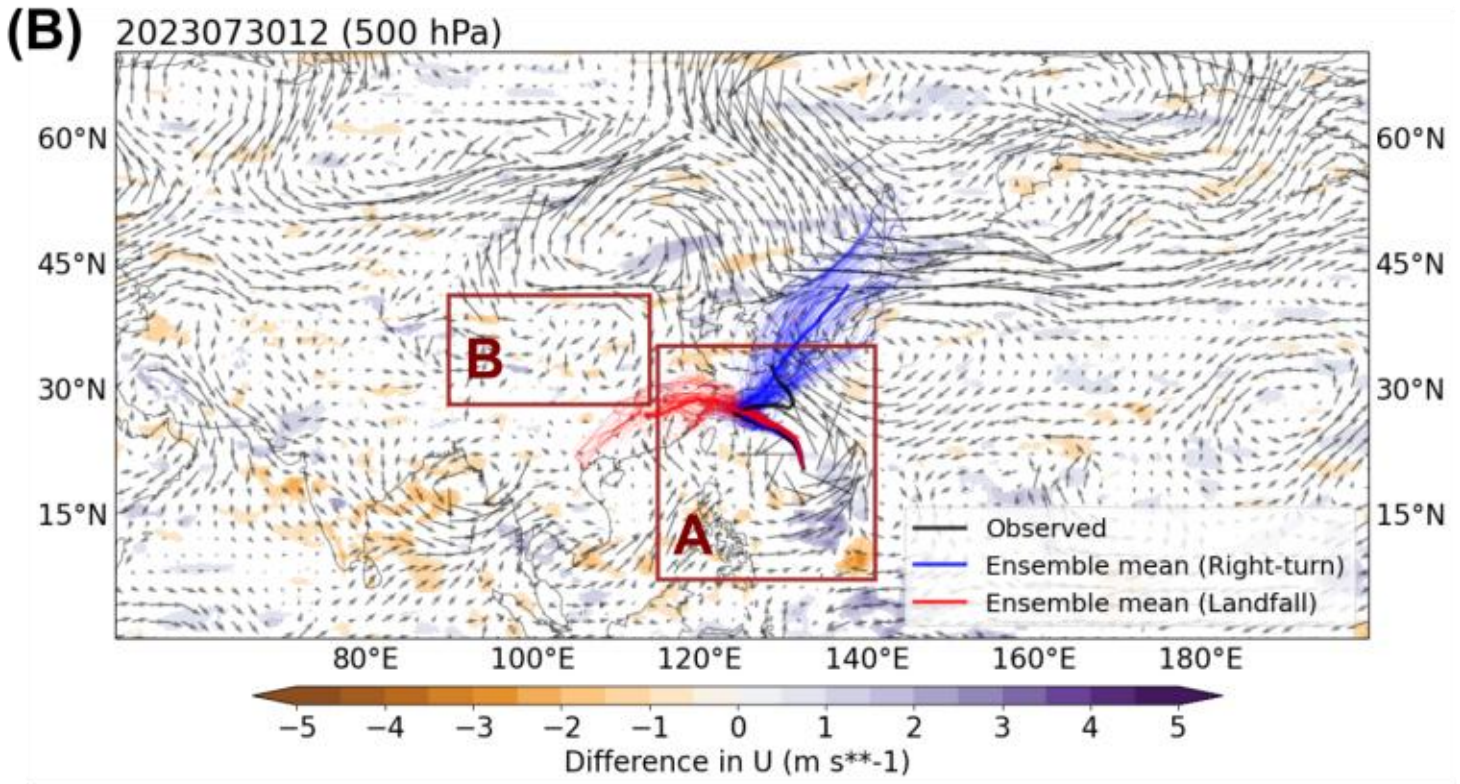
(B)
2023073012 (500 hPa)
B
A
Observed
Ensemble mean (Right-turn)
Ensemble mean (Landfall)
60°N
45°N
30°N
15°N
80°E
100°E
120°E
140°E
160°E
180°E
−5 −4 −3 −2 −1 0 1 2 3 4 5
Difference in U (m s**-1)

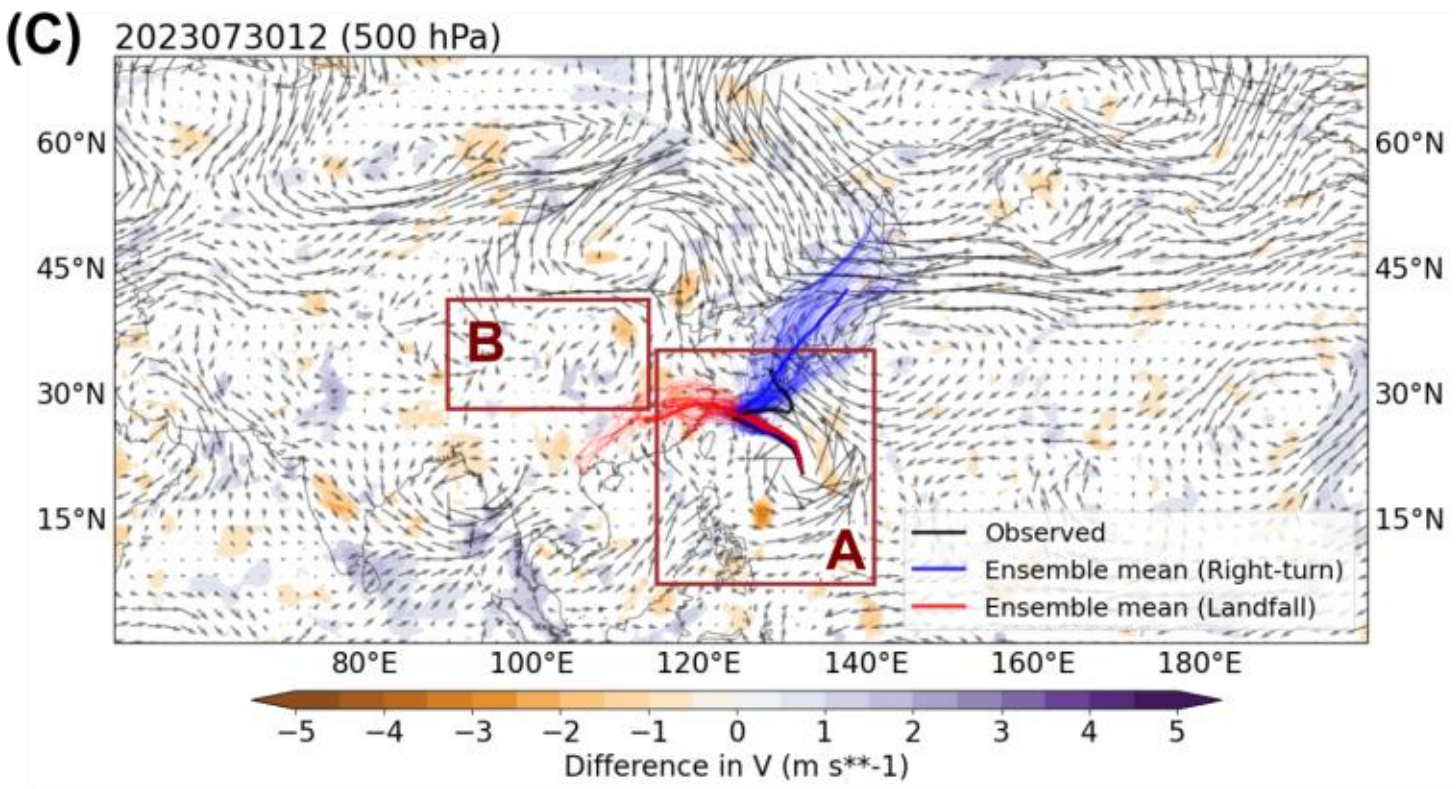
(C)
2023073012 (500 hPa)
B
A
Observed
Ensemble mean (Right-turn)
Ensemble mean (Landfall)
60°N
45°N
30°N
15°N
80°E
100°E
120°E
140°E
160°E
180°E
−5 −4 −3 −2 −1 0 1 2 3 4 5
Difference in V (m s**-1)

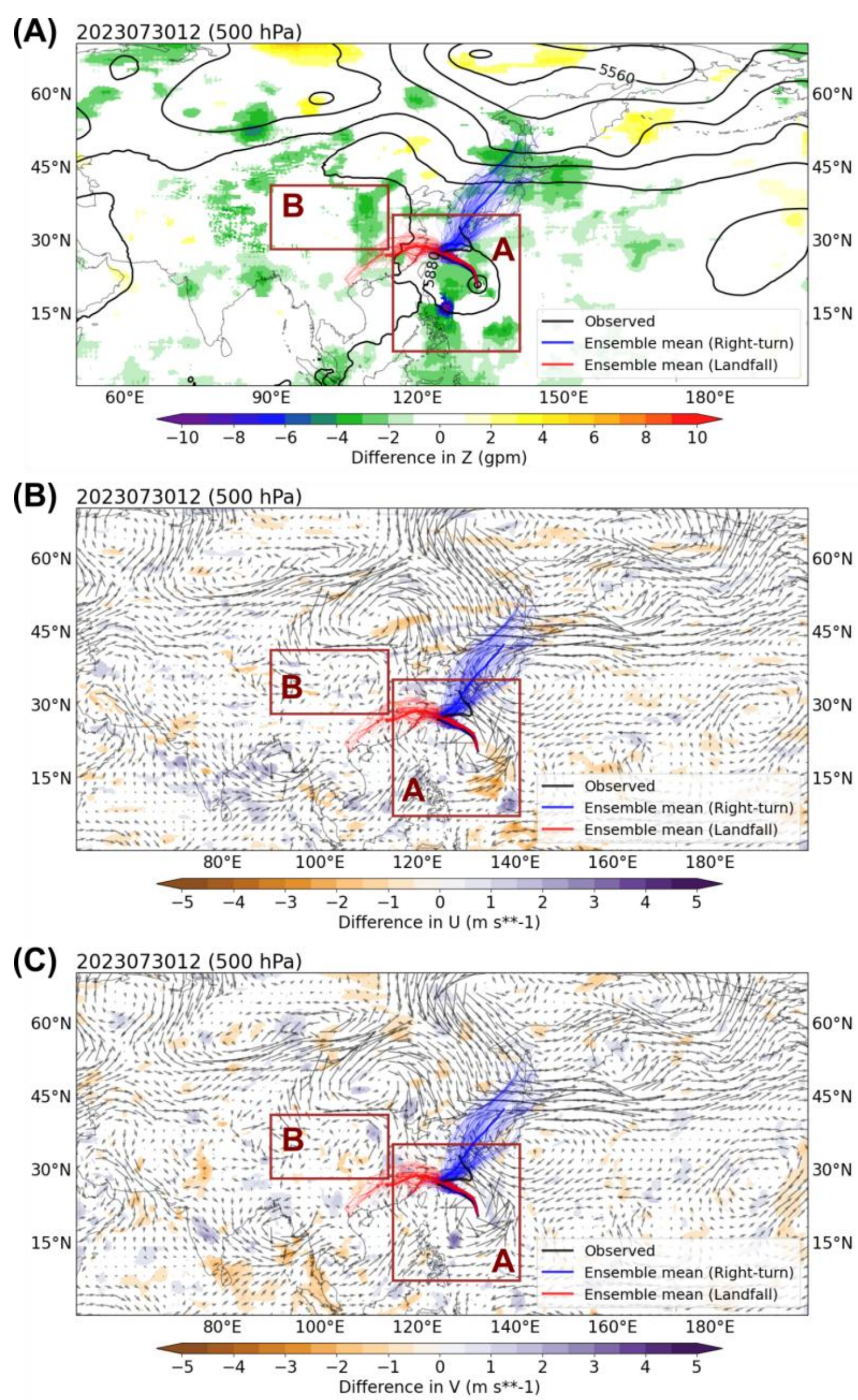

**Fig. S4. Differences in initial conditions between the right-turn and landfall clusters in Pangu-Weather ensemble experiments.** **(A)** Initial conditions of 500-hPa geopotential in Pangu-Weather control run initialized at 1200UTC 30th July 2023 (contour, unit: gpm), and their differences (shading, unit: gpm) between the right-turn (blue curves) and landfall (red curves) clusters in Pangu-Weather ensemble experiments. Only differences with p-values < 0.0001 were plotted. The two boxes denote the two key regions (Region A and B) affecting Khanun's predicted track. The black curve denotes the observed trajectory of Khanun. **(B to C)** Same as **(A)**, except for **(B)** 500-hPa zonal wind (unit: m/s) and **(C)** 500-hPa meridional wind (unit: m/s), respectively.

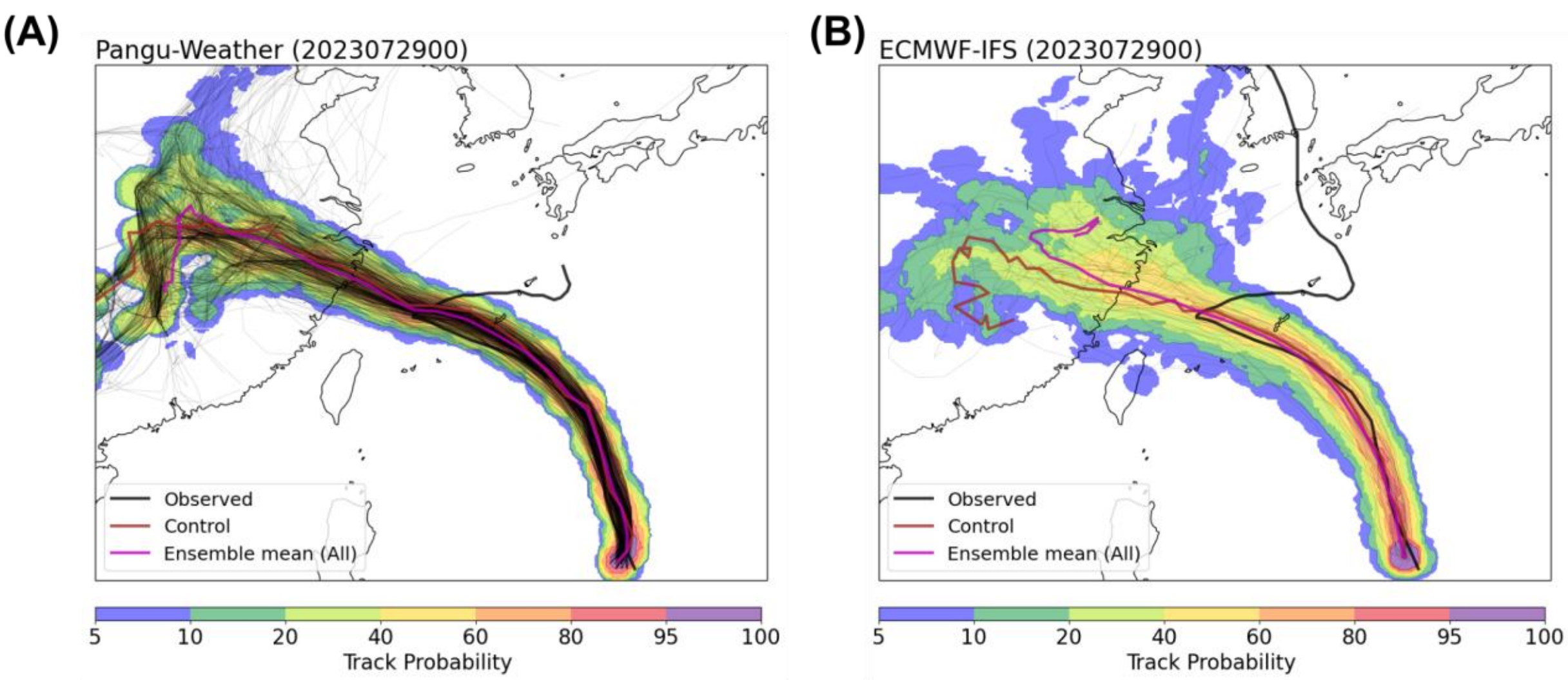

**Fig. S5. Smaller uncertainty of probabilistic prediction on Khanun's abrupt right turn at other initialization time. (A)** Same as Fig. 2A in the main text, except for results initialized at 1200UTC 29th July 2023. **(B)** Same as **(A)**, except for ensemble results of ECMWF-IFS.